\documentclass[conference, 10pt, twocolumn]{IEEEtran}

\usepackage[utf8]{inputenc} 
\usepackage[T1]{fontenc}
\usepackage{url}
\usepackage{ifthen}
\usepackage{cite}
\usepackage[cmex10]{amsmath} 
\usepackage{balance}
\usepackage{graphicx}
\usepackage{amssymb}
\usepackage{color}
\usepackage{algorithm}
\usepackage{algpseudocode}
\usepackage{xcolor}
\algnewcommand\algorithmicinput{\textbf{Input:}}
\algnewcommand\INPUT{\item[\algorithmicinput]}
\algnewcommand\algorithmicoutput{\textbf{Output:}}
\algnewcommand\OUTPUT{\item[\algorithmicoutput]}
\usepackage{mathtools}

\newcommand{\Fig}[1]{Fig.~\textup{\ref{#1}}}
\graphicspath{{figures/}}
\usepackage{subcaption}
\newtheorem{definition}{Definition}
\newtheorem{lemma}{Lemma}
\newtheorem{theorem}{Theorem}
\newtheorem{corollary}{Corollary}
\newtheorem{remark}{Remark}
\newtheorem{example}{Example}

\newcommand{\F}{\mathbb{F}}

\newcommand{\remove}[1]{}

\usepackage{tikz}
\usetikzlibrary{positioning}
\usepackage{pgfplots}

\def\supp{\qopname\relax{no}{supp}}

\newcommand\nc\newcommand
\nc\bfa{{\boldsymbol a}}\nc\bfA{{\bf A}}\nc\cA{{\mathcal A}}
\nc\bfb{{\boldsymbol b}}\nc\bfB{{\bf B}}\nc\cB{{\mathcal B}}

\nc\bfd{{\boldsymbol d}}\nc\bfD{{\bf D}}\nc\cD{{\mathcal D}}
\nc\bfe{{\boldsymbol e}}\nc\bfE{{\bf E}}\nc\cE{{\mathcal E}}
\nc\bff{{\boldsymbol f}}\nc\bfF{{\bf F}}\nc\cF{{\mathcal F}}
\nc\bfg{{\boldsymbol g}}\nc\bfG{{\bf G}}\nc\cG{{\mathcal G}}
\nc\bfh{{\boldsymbol h}}\nc\bfH{{\bf H}}\nc\cH{{\mathcal H}}
\nc\bfi{{\boldsymbol i}}\nc\bfI{{\bf I}}\nc\cI{{\mathcal I}}
\nc\bfj{{\boldsymbol j}}\nc\bfJ{{\bf J}}\nc\cJ{{\mathcal J}}
\nc\bfk{{\boldsymbol k}}\nc\bfK{{\bf K}}\nc\cK{{\mathcal K}}
\nc\bfl{{\boldsymbol l}}\nc\bfL{{\bf L}}\nc\cL{{\mathcal L}}
\nc\bfm{{\boldsymbol m}}\nc\bfM{{\bf M}}\nc\cM{{\mathcal M}}
\nc\bfn{{\boldsymbol n}}\nc\bfN{{\bf N}}\nc\cN{{\mathcal N}}
\nc\bfo{{\boldsymbol o}}\nc\bfO{{\bf O}}\nc\cO{{\mathcal O}}
\nc\bfp{{\boldsymbol p}}\nc\bfP{{\bf P}}\nc\cP{{\mathcal P}}
\nc\bfq{{\boldsymbol q}}\nc\bfQ{{\bf Q}}\nc\cQ{{\mathcal Q}}
\nc\bfs{{\boldsymbol s}}\nc\bfS{{\bf S}}\nc\cS{{\mathcal S}}
\nc\bft{{\boldsymbol t}}\nc\bfT{{\bf T}}\nc\cT{{\mathcal T}}
\nc\bfu{{\boldsymbol u}}\nc\bfU{{\bf U}}\nc\cU{{\mathcal U}}
\nc\bfv{{\boldsymbol v}}\nc\bfV{{\bf V}}\nc\cV{{\mathcal V}}
\nc\bfw{{\boldsymbol w}}\nc\bfW{{\bf W}}\nc\cW{{\mathcal W}}
\nc\bfx{{\boldsymbol x}}\nc\bfX{{\bf X}}\nc\cX{{\mathcal X}}
\nc\bfy{{\boldsymbol y}}\nc\bfY{{\bf Y}}\nc\cY{{\mathcal Y}}
\nc\bfz{{\boldsymbol z}}\nc\bfZ{{\bf Z}}\nc\cZ{{\mathcal Z}}
\nc\od{{\bar d}}\nc\ow{{\bar w}}\nc\odelta{{\bar\delta}}
\nc\ox{{\bar x}}\nc\oy{{\bar y}}\nc\ou{{\bar u}}
\nc\oh{{\bar h}}

\DeclareMathOperator{\wt}{wt}
\DeclareMathOperator{\rk}{rk}

\newcommand{\code}{\ensuremath{\mathcal{C}}}

\newcommand{\T}{\ensuremath{\mathcal{T}}}
\newcommand{\A}{\ensuremath{\mathcal{A}}}

\newif\ifcomment
\commentfalse 

\newif\ifcommentLater
\commentLaterfalse

\usepackage{flushend}
\flushend

\title{Secrecy and Accessibility in Distributed Storage}

\author{
  \IEEEauthorblockN{Lukas Holzbaur\IEEEauthorrefmark{1}, Stanislav Kruglik\IEEEauthorrefmark{2}, Alexey Frolov\IEEEauthorrefmark{2}, Antonia Wachter-Zeh\IEEEauthorrefmark{1}}
	
 \IEEEauthorblockA{\small \IEEEauthorrefmark{1} Institute for Communications Engineering, Technical University of Munich, Germany
    }
 \IEEEauthorblockA{\small \IEEEauthorrefmark{2} Skolkovo Institute of Science and Technology, Moscow, Russia
    }
  {\small lukas.holzbaur@tum.de, stanislav.kruglik@skoltech.ru, al.frolov@skoltech.ru, antonia.wachter-zeh@tum.de}
  \thanks{L. Holzbaur was supported by TU Munich -- Institute for Advanced Study, funded by the German Excellence Initiative and EU 7th Framework Programme under Grant Agreement No. 291763 and the German Research Foundation (Deutsche Forschungsgemeinschaft, DFG) under Grant No. {WA3907/1-1}. The work of S. Kruglik was supported by RFBR according to the research projects no. 19-01-00364 and 19-37-90022.}
}
\IEEEoverridecommandlockouts
\begin{document}

\maketitle

\begin{abstract}
A \emph{distributed storage system} (DSS) needs to be efficiently accessible and repairable. Recently, considerable effort has been made towards the latter, while the former is usually not considered, since a trivial solution exists in the form of systematic encoding. However, this is not a viable option when considering storage that has to be secure against eavesdroppers. This work investigates the problem of efficient access to data stored on an DSS under such security constraints. Further, we establish methods to balance the access load, i.e., ensure that each node is accessed equally often. We establish the capacity for the alphabet independent case and give an explicit code construction. For the alphabet-dependent case we give existence results based on a random coding argument.
\end{abstract}

\section{Introduction}

The problem of \emph{locality}, i.e., the ability of a distributed storage system (DSS) to recover a specified number of node failures from only a small number of other nodes, has been studied intensively in recent literature \cite{Balaji2018, Liu2018363}. This increased interest is driven by the desire to avoid large overhead and organizational complexity stemming from involving a large number of nodes in the repair process. By the same reasoning, it is not desirable for a system having to connect to a large number of nodes to recover data requested by a user. When user data is not secret, this is not an issue, as a systematic encoding of the data offers a simple and optimal solution, where a user retrieving data can obtain his files from a single server. However, when any number of $t$ nodes should not be able to learn anything about the user data, systematic encoding is not possible anymore. In this work, we consider this problem of \emph{local access} in secure DSS. While this problem is by its motivation closely related to locally repairable codes, we stress that we are \emph{not} concerned with secure storage codes that offer local repair of failed nodes, as, e.g., considered in \cite{Agarwal2016,Kadhe2017}.

Assume $k_D$ data files are given, which can be considered as symbols of some alphabet. We want to store these files on $n$, where $n>k_D$, servers such that: (a) a passive eavesdropper having access to up to $t$ servers cannot obtain any information; (b) each file can be recovered by accessing no more than $r$ servers (clearly, $r > t$); (c) the system can tolerate a given number of server failures without data loss.

\begin{figure}[!h]
\center
    \def\x{1}

\begin{tikzpicture}

\node (S1) at (0,0) [draw,thick,minimum width=\x*2.3cm,minimum height=\x*1.4cm,] {$\mathbf{X_0+Z}$};
\node (S2)  [right=\x*0.3cm of S1, draw,thick,minimum width=\x*2.3cm,minimum height=\x*1.4cm] {$\mathbf{X_1+Z}$};
\node (S3)  [right=\x*0.3cm of S2, draw,thick,minimum width=\x*2.3cm,minimum height=\x*1.4cm] {$\mathbf{X_0+X_1+Z}$};

\end{tikzpicture}
   \caption {Example of a secure DSS storing two files such that each file can be retrieved from two nodes.}
    \label{fig:ExDSS}
\end{figure}
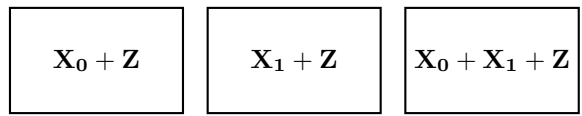

Let us show the requirements with use of simple example. Assume we have two files $X_0$ and $X_1$ and want to store them in a secure way on three servers. A possible scheme is shown in \Fig{fig:ExDSS}. Here, $Z$ denotes a random variable over the same alphabet as $X_0$ and $X_1$. Clearly, an eavesdropper with access to a single server does not gain any information about the files. At the same time, accessing two servers is sufficient to recover each of the files, e.g., to recover file $X_0$ it suffices to read $X_1+Z$ and $X_0+X_1+Z$ and calculate $X_0 = (X_0+X_1+Z) - (X_1+Z)$. Note that the goal of a secure LRC as considered in \cite{Agarwal2016,Kadhe2017} would be to locally recover $X_0+Z$, $X_1+Z$, or $X_0+X_1+Z$ (codeword symbols) instead of $X_0$ or $X_1$ (data/message symbols).

Previous works that consider this problem include \cite{Huang2017CodingFS, bruck,Huang2017SecureRS}. However, the solutions presented there impose a very specific \emph{access structure}, where accessing any part of a file always requires accessing the same subset of nodes, while other nodes never participate in the retrieval of a file.

This problem is a particular case of a \emph{secret sharing} problem. In secret sharing, the common secret is distributed among the participants. The participants possess so-called ``shares'', such that only allowed subsets of participants can recover the secret and all other coalitions can derive no additional information. The most popular and well-investigated secret sharing scheme is an $(n,s)$-threshold scheme proposed and investigated in \cite{Shamir, Bla79}. In such a scheme there are $n$ shares and any subset of at least $s$ participants can recover the secret, while any set of $<s$ share provides no information about the secret. In \cite{Shamir, Bla79} the solution of this problem is formulated in terms of polynomials. The connection to Reed--Solomon codes was established in \cite{McElieceSarwate}.  In \cite{Massey93minimalcodewords}, secret sharing schemes with use of linear codes were considered and the connection in between access structure and minimal weight codewords of the dual code was established.  

In this paper, we require a special access structure, i.e. any file can be recovered by accessing at most $r$ servers. This is different from the threshold secret sharing as (a) we do not require any $r$ servers to be able to recover the file; (b) we want to recover a single file rather that the whole secret ($k$ files).

There has also been considerable work on secret sharing schemes with multiple secrets and specific access structures \cite{jackson1993multisecret,Jackson1996, HSU}. The difference to our work is that there the requirement is that any set of users that is not specified as part of the access structure of a given secret is not allowed to learn anything about this secret. Our definition of \emph{accessibility} is a relaxed version of this definition of an access structure, in the sense that we allow any set of $>t$ nodes to recover any information of the file, and only require existence of recovery sets of a specified size for each file/secret.

Another related problem is that of \emph{private information retrieval} (PIR) as posed in \cite{Chor1995}. There, the user wants to retrieve a file from a DSS such that any set of $t$ servers does not learn the index of the desired file. However, while we allow the servers to learn the index of the retrieved file but not the file \emph{contents}, the servers in the PIR setting are not allowed to know the index, but do know the file contents.

\section{Preliminaries}
\subsection{Notations}
Let $A$ be an $a\times b$ matrix and $\mathcal{A} \subseteq \{0,...,b-1\}$ be a set of integers. We denote by $A|_{\mathcal{A}}$ the restriction of the matrix $A$ to the columns indexed by $\mathcal{A}$. Further, we denote by $\left\langle A \right\rangle$ and $\left\langle A \right\rangle_c$ the row span and column span of $A$, respectively.  We also use the notation $[n] \triangleq \{0,\dots,n-1\}$.We denote a linear code of length $n$, dimension $k$, and distance $d$ over $\F_q$ by $[n,k,d]_q$. If the distance and/or field size are not important, we simply write $[n,k]$.

\subsection{Problem Description}

We are interested in information-theoretic security and begin by formally stating the problem of $t$-collusion secure storage.
\begin{definition}[$t$-Collusion Secure Storage]\label{def:CollusionSecureIT}
  An $(n,k_D,d,t)$ DSS stores $k_D$ files $X=(X_0, ..., X_{k_D-1}) \in \F^{k_D}$ on $n$ nodes. We denote the information stored in the system by $Y=(Y_0, ..., Y_{n-1}) \in \F^n$. The encoding function is given by the randomized mapping
  \begin{equation*}
      f: \F^{k_D} \mapsto \F^{n},
  \end{equation*}
  where by \emph{randomized} we mean that $f$ is defined by conditional probabilities $\Pr(Y|X)$.
  
  In what follows we assume a uniform distribution on $X$. By $H(X)$ we denote the entropy of a discrete random variable (RV) and by $I(X;Y) = H(X) - H(X|Y)$ we denote the mutual information in between discrete RVs $X$ and $Y$. 

  Any set of $n-d+1$ nodes recovers all files, i.e., $\forall \ \mathcal{D} \subseteq [n]$ with $|\mathcal{D}| \geq n-d+1$ it holds that
  \begin{equation*}
      H(X|Y_{\mathcal{D}}) = 0 \ .
  \end{equation*}
  The system is $t$-collusion secure, i.e., $\forall \ \T \subset [n]$ with $|\T| \leq t$ it holds that
  \begin{align*}
    I(X,Y_\T) = 0 \ .
  \end{align*}
\end{definition}
In accordance with the literature on secret sharing \cite{Shamir, Bla79} we say a storage system is \emph{ideal} if $H(Y_i) = H(X_j) \ \forall i\in [n], j\in [k_D]$. In this work we only consider ideal storage systems.

This security model is the same as in \cite{Agarwal2016,Kadhe2017}, but while these works considered local repair of nodes, we require the property of \emph{local access}.
\begin{definition}[Access Complexity]\label{def:AccessComplexity}
A storage system as in Definition~\ref{def:CollusionSecureIT} is of \emph{access complexity} $r$ if $\forall i \in [k_D] \ \exists \A \subset [n]$ with $|\A| \leq r$ such that
    $H(X_i | Y_{\A}) = 0  $.
  We call the set $\A$ a recovery set of $i$ and say it is \emph{minimal} if
  \begin{equation*}
      H(X_i | Y_{\A \setminus a}) > 0 \ , \forall \ a \in \A \ .
  \end{equation*}
\end{definition}
Our goal is to maximize the secure storage rate 
\begin{equation}
    R \triangleq \frac{k_D}{n}
\end{equation}
of the DSS under the given constraint. We refer to the highest achievable rate as the \emph{capacity} of the system.

Definitions~\ref{def:CollusionSecureIT} and~\ref{def:AccessComplexity} define a $t$-collusion secure storage system in the general case, where the data stored at the nodes is an arbitrary function of the randomness and the files. To transform this problem into a code design task, we define the notion of a linear $t$-collusion secure DSS with linear access complexity. We will show in Section~\ref{sec:bounds} that for the alphabet-independent case it suffices to consider linear schemes to achieve the capacity, i.e., the highest possible storage rate.

\begin{definition}[$t$-Collusion Secure Code]\label{def:code}
  Let $\code$ be a linear $[n,k_D+k_S, d]$ code. We say that $\code$ is a \emph{$t$-collusion secure code} if there exists a matrix $G$ with $\left\langle G \right\rangle = \code$ that can be written as
\begin{equation} \label{eq:G}
  G = \left(
    \begin{array}{c}
      G_D\\
      G_S
    \end{array}\right) \ ,
\end{equation}
where $G_D \in \F^{k_D \times n}$ and $G_S \in \F^{k_S \times n}$, such that $d_S^\perp >t$, where $d_S^\perp$ is the distance of the code with parity check matrix~$G_S$.
\end{definition}

For a linear $t$-collusion secure storage system the access complexity can be directly related properties of the generator matrix.
\begin{definition}[Linear Access Complexity] \label{def:linearAccessComplexity}
For a code $\code$ as in the Definition~\ref{def:code} the \emph{access complexity} $r$ is given by
\begin{equation*}
  r = \max_{i \in [k_D]} \min_{\substack{\mathcal{A}\subseteq [n] \\ e_i \in \left\langle G|_{\mathcal{A}} \right\rangle_c }} |\mathcal{A}| \ ,
\end{equation*}
where $e_i$ denotes the $i$-th unit vector.
\end{definition}

If a $t$-collusion secure code is of access complexity $r$ by Definition~\ref{def:linearAccessComplexity}, each of the $k_D$ files can be obtained from the stored symbols by performing the respective linear combinations that result in the unit vector when performed on the columns of $G$. 

\begin{theorem} \label{th::secrecy}
  A $t$-collusion secure code as in Definition~\ref{def:code} gives a $t=(d_S^\perp -1)$-collusion secure storage system as in Definition~\ref{def:CollusionSecureIT}.
\end{theorem}
\begin{IEEEproof}
  The word $Y$ can be split into two parts by
  \begin{align*}
    Y &= (X_0,...,X_{k_D-1}, Z_0,...,Z_{k_s-1}) \cdot G \\
    &= \underbrace{(X_0,...,X_{k_D-1})\cdot G_{D}}_{Y_D} + \underbrace{(Z_0,...,Z_{k_s-1})\cdot G_S}_{Y_S} \ .
  \end{align*}
  Any $t = d_S^\perp-1$ columns of $G_S$ are linearly independent. Therefore, for any set of nodes $\mathcal{T}\subset [n]$ with $|\mathcal{T}| \leq t$ choosing $Z_0,...,Z_{k_s-1}$ i.i.d. random gives $Y_{S,\mathcal{T}} \in \F^{1\times |\mathcal{T}|}$ distributed uniformly over the vectors of length $t$ and therefore independent of $X$. It follows that
  \begin{equation*}
    I(X,Y_{\mathcal{T}}) = I(X,Y_{D,\mathcal{T}} + Y_{S,\mathcal{T}}) = 0 \ .
  \end{equation*}
\end{IEEEproof}

\begin{remark}
  Trivially, it holds that $r > t$ and by the Singleton bound it holds that $t \leq k_S$.
\end{remark}

\section{Bounds} \label{sec:bounds}

This section is devoted to the bounds on the distance $d$ and storage dimension $k_D$ for the case of general $t$-collusion secure storage system, i.e. we consider an arbitrary randomized mapping $f$. We note that the bounds do not depend on $r$ and are based on the techniques from \cite{Agarwal2016}. 

\begin{definition}
Define the support of the map $f$ by
\begin{equation*}
    \supp(f(X))=\{Y : \Pr(Y|X)\ne 0\} \ .
\end{equation*}
\end{definition}

\begin{theorem}\label{thm:DistanceNonLinear}
Consider a $t$-collusion secure $(n,k_D,d,t)$ DSS. The following holds for the parameters of the system
  \begin{equation*}
    d \leq d_q^{*}(n-t,q^{k_D}) \ ,
    \end{equation*}
where $d_q^{*}(n,M)$ is any upper bound on the distance of a $q$-ary code of length $n$ and of size $M$.
\end{theorem}
\begin{IEEEproof}
  Let us fix the set $\T \subseteq[n]$ with $|\T| = t$, $X\in \F_q^{k_D}$ and some $Y \in \supp(f(X))$. Due to the fact that $I(X,Y_\T)=0$ we claim that for any $X' \ne X$ there must exist $Y'\in \supp(f(X'))$ with $Y'_\T=Y_\T$.
   
  Let $\code_0 = \{Y'\in \supp(f(X')): X' \in \F_q^{k_D}, Y'_\T=Y_\T\}$
  such that $|\code_0 \cap \supp(f(X'))|=1$. We note that $|\code_0| = q^{k_D}$. Since $\code_0$ has fixed values on the coordinates $\T \subseteq[n]$, the code $\code_{[n]\setminus \T} \in \F_q^{n-t}$ must be a $q$-ary code of length $n-t$ and size $q^{k_{D}}$ and theorem statement follows.
\end{IEEEproof}

\begin{corollary}\label{cor::upper_bound}
We note that the following inequality holds
$$k_D \leq \log_q M_q^*(n-t,d),$$
where $M_q^*(n,d)$ is any upper bound on the size of a code with length $n$ and distance $d$.
\end{corollary}

\begin{theorem}[Capacity of $t$-Collusion Secure Storage] \label{thm:capacity}
  The capacity of an ideal $(n,k_D,d,t)$ DSS with access complexity $r=t+1$ is
  \begin{equation*}
      C_{(n,k_d,d,t)} = \frac{k_D}{k_D+t} \ .
  \end{equation*}
\end{theorem}
\begin{IEEEproof}
  \textbf{Converse:} By the Singleton bound and Theorem~\ref{thm:DistanceNonLinear} it holds that
      $1 \leq d \leq d_q^{*}(n-t,q^{k_D}) \leq n-t-k_D+1$
 and thus $n \geq k_D+t$.
 
  \textbf{Achievability:} Given by generalized Reed--Solomon (GRS) codes, as shown later in Lemma~\ref{lem:GRSoptimal}, as they achieve both the upper bound on the storage rate and the lower bound of $r > t$ on the access complexity.
\end{IEEEproof}
We refer to a $t$-collusion secure code or DSS as \emph{optimal} if it attains this capacity.

\section{Constructions}\label{sec:constructions}

In the following, we will determine a sufficient and necessary condition for a code to be a $t$-collusion secure code. The result presented in Theorem~\ref{thm:codeInfoSet} is similar to that in \cite{Huang2017CodingFS}. However we take a different approach, which will enable us to further extend these results in the following. 

\begin{lemma}\label{lem:matrixFullRank}
Let $A_1 \in \F^{a \times b}$ and $A_2 \in \F^{a \times b}$. Then $\rk(A_1\cdot A_2^T) = a$ if and only if
    $\left\langle A_1 \right\rangle^\perp \cap \left\langle A_2 \right\rangle = \{0\} .$
\end{lemma}
\begin{IEEEproof}
Assume $\rk(A_1 \cdot A_2^T) < a$. Then there exists a non-zero $v \in \F^{a\times 1}$ such that $A_1 \cdot A_2^T v = A_1 \cdot (v A_2)^T = 0$. It follows that $vA_2 \in  \left\langle A_1 \right\rangle^\perp$. Clearly $vA_2 \in  \left\langle A_2 \right\rangle $ and the lemma statement follows.
\end{IEEEproof}

\begin{theorem}\label{thm:codeInfoSet}
An $[n,k,d]$ code $\code$ is an optimal $(n,k_D=k-t,d,t)$ secure storage code of access complexity $r=t+1$ if and only if it contains an $[n,t]$ MDS subcode.
\end{theorem}
\begin{IEEEproof}
The necessity follows directly from Theorem~\ref{thm:codeInfoSet} and the Singleton bound. To show that it is also sufficient for a code to contain an $[n,t]$ MDS subcode, consider a code $\code$ with a generator matrix $G$ as in Definition~\ref{def:code}, where $G_S \in \F^{t\times n}$ is a generator matrix of an $[n,t]$ MDS code $\code_S$. Then the generator matrix of $\code$ can be written as
\begin{equation*}
   G = \left( \begin{smallmatrix}
   M & \mathbf{0} \\
   \mathbf{0} & I_t
   \end{smallmatrix}\right) \cdot
   \left(\begin{smallmatrix}
   G_D' \\ G_S
   \end{smallmatrix}\right) \ ,
\end{equation*}
 where $G_D' \in \F^{k_D \times n}$, $M \in \F^{k_D \times k_D}$ is of full rank, and $I_t \in \F^{t \times t}$ is an identity matrix. Let $B\in \F^{k_D \times n}$ be a matrix containing codewords of weight $t+1$ from $\code_S^\perp$ as rows, i.e., $G_S \cdot B^T = 0$.
 Define $M = (G_D' \cdot B^T)^{-1}$, then
\begin{align}  \label{eq:recoveryRS}
 G \cdot B^T &=
 \left(\!\begin{smallmatrix}
   M\cdot G_D' \cdot B^T \\ G_S \cdot B^T
 \end{smallmatrix}\!\right) \!=\!
 \left(\!\begin{smallmatrix}
   (G_D' \cdot B^T)^{-1} \cdot  G_D' \cdot B^T \\ \mathbf{0}
 \end{smallmatrix}\!\right)
 \!=\!
 \left(\!\begin{smallmatrix}
   I_{k_D} \\ \mathbf{0}
 \end{smallmatrix}\!\right)
\end{align}
and it follows that each of the $k_D$ unit vectors can be obtained as a linear combination of exactly $r=t+1$ columns, as required by Definition~\ref{def:linearAccessComplexity}.

It remains show that $G_D' \cdot B^T$ is invertible, i.e., of full rank. By Lemma~\ref{lem:matrixFullRank} this is the case if and only if 
\begin{equation}\label{eq:necessarySpanEmpty}
    \left\langle G_D' \right\rangle^\perp \cap \left\langle B \right\rangle = \{0\} \ .
\end{equation}
\remove{We proof the existence of such a matrix $B$.} Without loss of generality assume that the set $\{0,1,...,k_D+t-1\}$ is an information set of the code $\code$. Let the rows of $B$ be codewords of $\code_S^\perp$, where the support of the $i$-th row is given by $\supp(B_{i,:}) = \{i,k_D,k_D+1,...,k_D+t-1\}$. The existence of such codewords in the code $\code_S^\perp$ follows from observing that $G_S$ is the parity check matrix of an MDS code of distance $d_S^\perp =t+1$. Such a code contains a codeword with $\supp(c) = \mathcal{S}$ for any $\mathcal{S} \subset [n]$ with $|\mathcal{S}| = d_S^\perp$ (cf. \cite[Chapter~11]{MacWilliams1977}). It follows that the indices of the non-zero columns of $B$ are given by $\supp (B) = \{0,1,...,k_D+t-1\}$. Denote by $\overline{G}_D'$, $\overline{B}$, and $\overline{G}_S$ the respective matrices restricted to the columns indexed by $\supp(B)$. Note that $G_D' \cdot B^T = \overline{G}_D' \cdot \overline{B}^T$. Now assume on the contrary that this matrix is not of full rank, i.e., there exists a non-zero vector $v\in \F^{1 \times k_D}$ such that $(\overline{G}_D' \cdot \overline{B}^T) \cdot v^T = 0$. Since $G_S \cdot B^T \cdot v^T = \overline{G}_S \cdot \overline{B}^T \cdot v^T=0 $ by definition, these conditions give
\begin{align*}
 \underbrace{\left(\begin{smallmatrix}
   \overline{G}_D' \\ \overline{G}_S
 \end{smallmatrix}\right)}_{\overline{G}'}
 \cdot (v \cdot \overline{B})^T = \mathbf{0} \ .
\end{align*}
The support of $B$ is an information set of the code $\code$ by definition, so the matrix $\overline{G}' \in \F^{k_D+t \times k_D+t}$ is of full rank $k_D+t$, hence the equation can only be fulfilled if $v \cdot \overline{B} = 0$. This is a contradiction since the rows of $B$ are clearly independent. It follows that $G_D' \cdot B^T$ is of full rank.
\end{IEEEproof}

\begin{lemma}\label{lem:GRSoptimal}
An $[n,k=k_D+t]$ generalized Reed-Solomon (GRS) code is an optimal $t$-collusion secure code for any $t$ with $0 < t < k_D$.
\end{lemma}
\begin{IEEEproof}
It is easy to see that a GRS code contains an $[n,t]$ MDS subcode for any $t$ in the given range 
and the optimality follows from Theorem~\ref{thm:codeInfoSet}.
\end{IEEEproof}

\begin{example}\label{eq:constructionRS}
    We construct an $(n=6,k_D=2,d=3,t=2)$ secure storage code over $\F_7$ of access complexity $r=t+1=3$. We begin with a $[6,4]$ GRS code with generator matrix in Vandermonde form, where the first two rows give $G_D$ and the last two rows give $G_S$, i.e.,
  \begin{align*}
      G_D' = \left(\begin{smallmatrix}
      1&1&1&1&1&1\\
      1&2&3&4&5&6\\
      \end{smallmatrix} \right)
  \;\; \text{and} \;\;
      G_S = \left(\begin{smallmatrix}
      1&4&2&2&4&1\\
      1&1&6&1&6&6\\
      \end{smallmatrix}\right) \ .
  \end{align*}
  For $B$ we use 
  \begin{equation*}
      B = \left( \begin{smallmatrix}
      1 & 0 & 2 & 1 & 0 & 0 \\
        0 & 1 & 3 & 2 & 0 & 0
      \end{smallmatrix} \right)
  \end{equation*}
  and obtain 
      $M = (G_D' \cdot B^T)^{-1} = \left(\begin{smallmatrix}
      4&5\\
      1&6
      \end{smallmatrix}\right) $.
  Finally, we get 
  \begin{align*}
      G = \left(\begin{smallmatrix}
      4&5&0&0\\
      1&6&0&0\\
      0&0&1&0\\
      0&0&0&1
      \end{smallmatrix}\right) \cdot \begin{pmatrix}
      G_D' \\ G_S \end{pmatrix}
      = 
     \left( \begin{smallmatrix}
      2 & 0 & 5 & 3 & 1 & 6 \\
    0 & 6 & 5 & 4 & 3 & 2 \\
    1 & 4 & 2 & 2 & 4 & 1 \\
    1 & 1 & 6 & 1 & 6 & 6
    \end{smallmatrix} \right) \ .
  \end{align*}
  It is easy to check that \eqref{eq:recoveryRS} holds.
  
\end{example}

\section{Load Balancing}\label{sec:loadBalancing}
While Theorem~\ref{thm:codeInfoSet} provides a simple sufficient and necessary condition for a code to be an optimal $t$-collusion secure code, the access structure used for proving the optimality, which is similar to that in \cite{Huang2017CodingFS}, has some disadvantages. As the support of $B$ is restricted to a specific set of $k_D+t$ positions, where $k_D$ columns contain a $k_D\times k_D$ identity matrix, the access to the nodes would be very unbalanced, as there are $t$ nodes which are accessed in the recovery of \emph{any} file, $k_D$ which are accessed in the recovery of exactly one file each, and $n-(k_D+t)$ nodes which are \emph{never} accessed. In this section, we consider constructions with more balanced access structures. We begin by formally defining our objective.
\begin{definition}[Balanced Access]\label{def:balancedB}
  Let $B \in \F^{k_D \times n}$ be a matrix where every row has weight $t+1$. We say the matrix is balanced if every column is of weight 
  \begin{equation*}
  \left\lfloor \frac{k_D(t+1)}{n} \right\rfloor \leq \mathrm{wt} (B_{:,j}') \leq \left\lceil \frac{k_D(t+1)}{n} \right\rceil, \ \forall \ j \in [n] \ .
  \end{equation*}
\end{definition}

\begin{example}
    Consider the same system as in Example~\ref{eq:constructionRS}. The matrix $B$ given there is clearly not balanced, as $Y_2$ and $Y_3$ are both required for the retrieval of $X_0$ and $X_1$. By Defintion~\ref{def:balancedB} the weight of each column in a balanced access structure $B'$ is $\wt(B_{:,j}') = k_D(t+1)/n = 1$. It is easy to check that
    \begin{equation*}
        B' = \left(\begin{smallmatrix}
    0 & 1 & 2 & 0 & 0 & 6 \\
    1 & 0 & 0 & 5 & 6 & 0
    \end{smallmatrix}\right)
    \end{equation*}
    is balanced as in Definition~\ref{def:balancedB} and satisfy condition that $G_D' \cdot B^T$ is invertible. The respective generator matrix is given by
    \begin{equation*}
        G = \left(\begin{smallmatrix}
      2&2&0&0\\
      5&2&0&0\\
      0&0&1&0\\
      0&0&0&1
      \end{smallmatrix}\right) \cdot \begin{pmatrix}
      G_D' \\ G_S \end{pmatrix}
      = 
     \left( \begin{smallmatrix}
    4 & 6 & 1 & 3 & 5 & 0 \\
    0 & 2 & 4 & 6 & 1 & 3 \\
    1 & 4 & 2 & 2 & 4 & 1 \\
    1 & 1 & 6 & 1 & 6 & 6
    \end{smallmatrix} \right) \ .
    \end{equation*}
\end{example}

In what follows we use the language of parity-check matrices to present our results. In accordance to (\ref{eq:G}) we have $\code^\perp = \code_D^\perp \cap \code_S^\perp$ and thus 
\begin{equation} \label{eq:HS}
{H}_D = \left( \begin{array}{c} A \\ {H} \end{array} \right), \qquad {H}_S = \left( \begin{array}{c} {H} \\ B' \end{array} \right),
\end{equation}
where $H_D \in \F_q^{(n-k_D) \times n}$, $H_S \in \F_q^{(n-k_S) \times n}$ and $H \in \F_q^{(n-k_D-k_S) \times n}$ are parity-check matrices of codes $\code_D$, $\code_S$ and $\code$ accordingly. For $A \in \F_q^{k_S \times n}$, $B' \in \F_q^{k_D \times n}$, the matrix 
\begin{equation*}
\left( \begin{smallmatrix} A \\ {H} \\ B' \end{smallmatrix} \right)
\end{equation*}
forms a basis of $\F_q^n$ (the rows are linearly independent).

We note that $B'$ from (\ref{eq:HS}) can be used as an access structure. Indeed $\code_D^\perp = \left\langle H_D \right\rangle$ and thus $\code_D^\perp \cap \left\langle B' \right\rangle = \{0\}$ due to the linear independence. The next lemma describes all possible access structures.
\begin{lemma}\label{lemmaA1A2}
Let $H_S$ be given by (\ref{eq:HS}), then any valid access structure $B$ is given by
\begin{equation*}
    B = (A_1, A_2) \cdot H_S \ ,
\end{equation*}
where $A_1 \in \F^{k_D \times (n-k_D-k_S)}$, $A_2 \in \F^{k_d \times k_D}$, and $\rk(A_2) = k_D$.
\end{lemma}

\begin{IEEEproof}
Clearly $B \subset \code^\perp_S$, since we require $G_S \cdot B^T = 0$ by definition, so $B = A \cdot H_S$ for some $A\in \F^{k_D \times (n-k_S)}$. Now assume $\rk(A_2)<k_D$, i.e., there exists a non-zero vector $e$ such that $e\cdot A_2 = 0$. Then
\begin{align*}
   e \cdot B &= e\cdot (A_1, A_2) \cdot H_S \\
   &= (e \cdot A_1 \cdot H + \underbrace{e \cdot A_2 \cdot B'}_{=0})  \ .
\end{align*}
Thus, $e \in \code_D^\perp$ as $\code^\perp \subset \code_D^\perp$. This contradicts the condition $\code_D^\perp \cap \left\langle B \right\rangle = \{0\}$ (recall we need this condition for the matrix $M$ to be invertible) and we conclude that $\rk(A_2)=k_D$.
\end{IEEEproof}

We now consider two code constructions that have the load balancing property. We use the notation of the Vandermonde matrix

\begin{equation*}
  V_{i}^j \triangleq
  \left(\begin{smallmatrix}
    1 & \alpha^i & \cdots & \alpha^{(n-1)i} \\
    1 & \alpha^{i+1} & \cdots & \alpha^{(n-1)(i+1)} \\
    \vdots & \vdots && \vdots \\
    1 & \alpha^{j} & \cdots & \alpha^{(n-1)j} \\
  \end{smallmatrix} \right). 
\end{equation*}

Let $H = V_0^{n-k_D-t-1}$, $H_S = V_0^{n-t-1} = \left( \begin{array}{c} {H} \\ V_{n-k_D-t}^{n-t-1} \end{array}\right)$, $H_D = \left( \begin{array}{c} V_{n-t}^{n-1} \\ {H} \end{array}\right)$ and $k_S = t$ for the rest of the section.

\paragraph{Construction 1} Let us choose 
\[
B = (\mathbf{0}, A_2) \cdot H_S = A_2 V_{n-k_D-t}^{n-t-1},
\]
where the matrix $A_2$ is of full rank and chooses vectors of weight $n-k_D+1$ from $\left\langle V_{n-k_D-t}^{n-t-1} \right\rangle$. This is always possible as $\left\langle V_{n-k_D-t}^{n-t-1} \right\rangle$ forms an $[n, k_D, n-k_D+1]$ GRS code. The code $\code$ is optimal (see Theorem~\ref{thm:DistanceNonLinear}), i.e. $d = n-t-k_D+1$ and can have arbitrary access structure with $r = n-k_D+1$. This constructions works for any $n$, $k_D$ and $t$ but interesting for the case $r < k_D+t$ (as having $k_D+t$ symbols we can always reconstruct the whole codeword and thus all information symbols), which holds for $R = k_D/n > 0.5( 1 - (t-1)/n )$.

\paragraph{Construction 2}  Let $\alpha$ be a primitive element of $\F_q$, $n = q-1$, $\frac{n}{n-t}=a$, where $a$ is an integer, and $k_D = a$. We note that 
\[
(1 1 \ldots 1) H_S = v = (\underbrace{** \ldots *}_{a} 0 \underbrace{** \ldots *}_{a} 0 \ldots \underbrace{** \ldots *}_{a}),
\]
as $\sum\nolimits_{l=0}^{n-t-1} \alpha^{a j l} = 0$ for $j = 1, \ldots, n-t-1$. Thus the weight $v$ is exactly $t+1$. 

We mention that the code $\code_S^\perp$ is cyclic and thus contains all cyclic shifts of $v$. The $j$-th cyclic shift (to the right) of $v$ can be obtain as $a_j H_S$, where $a_j = (1 \: \alpha^{-j} \: \alpha^{-2j} \ldots \alpha^{-(n-t-1)j})$. Let us form the matrix $A$ with the columns $a_j$ and choose $B = (A_1 A_2) H_S$. Clearly, $B$ contains $k_D = a$ vectors of weight $t+1$ and $A_2$ has full rank as it is the Vandermonde matrix. Thus, our construction has the following parameters: $n=q-1$, $k_D = a$, $a | n$, $t = \frac{a-1}{a} n$, $r = t+1$, $d = n/a-a+1$. Clearly, the rate $R = k_D/n \leq 0.5 $ for this construction.

\section{Small Fields}

In Section~\ref{sec:constructions}, we showed how to find an optimal $t$-collusion secure storage code over $\F_q$ with optimal access complexity $r=t+1$ for all parameters $(n,k,d,t)$ whenever $n\geq q$. This construction relies on the code having an MDS subcode, which is (except for the trivial cases) only possible if q is comparable to the code length. This section deals with constructions small field size $q$, where we cannot use the MDS property.

\begin{lemma}
    The $m$-variate Reed--Muller (RM) code $\mathcal{R}(v,m)$ of order $v$ is a $t$-collusion secure storage code with length $n=2^m$, dimension $k=\sum_{i=0}^{v} \binom{m}{i}$, distance $d=2^{m-v}$, storage dimension $k_D = \binom{m}{v}$, collusion resistance $t=2^{v}-1$, and access complexity $r=2^v$.
 \end{lemma}
 \begin{IEEEproof}
 The length, dimension, and distance are the parameters of the RM code. For the storage, we use the $\binom{m}{v}$ highest order terms, i.e., the rows in the generator matrix corresponding to functions of degree exactly $v$. There are $k_D = \binom{m}{v}$ such rows in any generator matrix of an $\mathcal{R}(v,m)$ code. The remaining rows give the matrix $G_S$. It is well-known that this is an $\mathcal{R}(v-1,m)$ code and that its dual is therefore an $\mathcal{R}(m-(v-1)-1,m)$ code of distance $d_S^\perp = t+1 = 2^{v}$ (cf. \cite[Chapter~13]{MacWilliams1977}). The access complexity follows from the properties of RM codes used in the Reed decoding algorithm (cf. \cite[Ch. 3, Theorem 14]{MacWilliams1977}). 
\end{IEEEproof}
Note that RM codes have $2^{m-v}$ disjoint recovery sets for each file, which can be used to obtain a balanced access structure.
 
 \begin{example}
 Let $\mathcal{R}(v,m)$ be an $m$-variate binary RM code of order $v$. Let $m=4$ and $v=2$. The parameters of this code are length $n=2^m= 16$, dimension $k=1+\binom{m}{1}+\binom{m}{2}$, and distance $d=2^{m-v}=4$.
 Consider the non-systematic encoder given in \cite[Chapter~13, 6]{MacWilliams1977} which can recover each of the symbols $a_{12},...,a_{34}$ corresponding to the second order terms from $r=4$ symbols. Let these be our data symbols, so $k_D = 6$. The remaining $k_S=5$ symbols give a first-order RM code $\mathcal{R}(v_S=1,4)$. The dual of this code is a $\mathcal{R}(m-v_s-1,m)$ code and therefore of distance $d_S^\perp = 4$. Hence the code is a $t=3$-collusion secure code storing $k_D=6$ files (bits) with distance $d=4$.
 
 With the Gilbert-Varshamov bound (cf. \cite{MacWilliams1977}) and Theorem~\ref{thm:DistanceNonLinear} we obtain
     $d \leq d_2^*(n-t,k_D) = d_2^*(13,6) = 4 $ ,
 for these parameters. Since $r=t+1$, the code is optimal for the given parameters $n,k,t,$ and $q$. However, note that $d_2^*(13,8) = 4$, so the code is not optimal w.r.t. $k_D$. 
 \end{example}
 
\section{Random Coding Bound}

In this section, we derive an asymptotic achievability bound by means of random coding.

\begin{theorem}
\label{th::random}
For any field size $q$ and sufficiently large $n$ there exists a $t = \tau n$ collusion secure storage code with distance $d = \delta n$, access complexity $r = n h_q(\tau) + o(n)$ and rate
\[
R = \frac{k_D}{n} \geq R^* = 1 - h_q(\delta) -  h_q(\tau) - o(1),
\]
where $h_q(x) = -x \log_q(x) - (1-x) \log_q (1-x) + x \log_q (q-1)$ denotes the $q$-ary entropy function.
\end{theorem}

\begin{IEEEproof}
Consider the parity-check matrix ${H}_S$ (see (\ref{eq:HS})). Recall that $B'$ can be used as an access structure and we only need to make it sparse. For this purpose, we transform the matrix ${H}_S$ to the form
\[
H'_S = \left( \begin{smallmatrix} I_{n-k_D-k_S} & \tilde{H} \\ \mathbf{0} & I_{k_D} \:\: \tilde{B} \end{smallmatrix} \right).
\]

Note that $B = \left( \mathbf{0} \:\: I_{k_D} \:\: \tilde{B} \right) = (A_1 A_2) H_S$ and $A_2$ has full rank. Thus in accordance to Lemma~\ref{lemmaA1A2} we constructed a code with access complexity $k_S + 1$. This approach can be applied to any matrix $H_S$.

In what follows we apply the random coding technique to obtain the values of $k_D$ and $k_S$. Consider the ensemble $\mathcal{E}_1$ of all parity-check matrices $H_S$ of size $(n-k_S) \times n$. The ensemble is defined as follows: positions in $H_S$ are i.i.d. random variables with uniform distribution over the set $[q]$. Let $m = n-k_S$. We can estimate the following probabilities
\begin{eqnarray*}
p_0 \triangleq \Pr\left( \rk(H_S) < m \right) = 1 - \prod_{i=0}^{m-1} (1 - 1/q^{n-i}) 
\leq \frac{m}{q^{n-m+1}} 
\end{eqnarray*}
and
$p_1 \triangleq \Pr\left( d(\mathcal{C}(H)) < \delta n\right) \leq q^{h_q(\delta) - 1 + \frac{k_D+k_S}{n} + o(1)}. $

The distance of the dual code can be estimated in exactly the same way but we need the ensemble $\mathcal{E}_2$ of the generator matrices $G_S$. It holds that
\begin{eqnarray*}
p_2 \triangleq \Pr\left( \rk(G_S) < k_S \right) \leq \frac{k_S}{q^{n-k_S+1}} 
\end{eqnarray*}
and
$p_3 \triangleq \Pr\left( d(\mathcal{C^\perp_S}) < \tau n\right) \leq q^{h_q(\tau) - \frac{k_S}{n} + o(1)}$.
Taking into account that each code $\mathcal{C}_S$ can be represented by the same number of parity-check (or generator) matrices, we conclude that the fraction of ``bad'' codes is less or equal to $p_0+p_1+p_2+p_3$. To conclude note that the fraction of ``bad'' codes can be made arbitrarily small (by increasing $n$) when $k_S = n h_q(\tau) + o(n)$ and $R < R^*$.
\end{IEEEproof}

Substitution of the first MRRW bound \cite{MRRW} for $k_q^*(n,d)$ from Corollary~\ref{cor::upper_bound} for $t=\tau n$ and $q=2$ gives the following asymptotic form of the upper bound 
\begin{equation}\label{upper_lp}
    R \leq (1-\tau)h_2\bigg(\frac{1}{2}-\sqrt{\frac{\delta}{1-\tau}(1-\frac{\delta}{1-\tau})}\:\bigg)+o(1)\ .
\end{equation}

\begin{figure}
\center
   \input{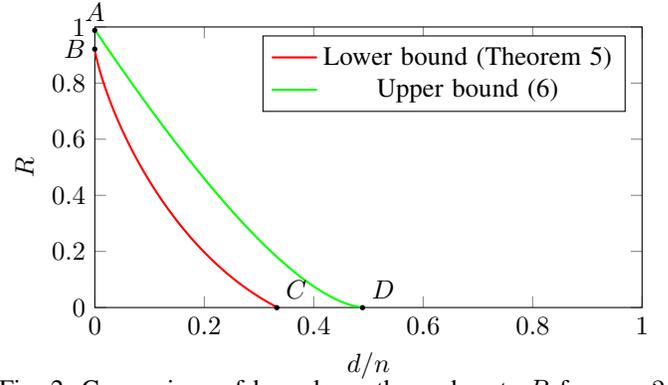}
   \vspace{-10pt}
   \caption {Comparison of bounds on the code rate $R$ for $q=2$ and $\tau=10^{-2}$. $A=(0,1-\tau),$ $B=(0,1-h_2(\tau)),$ $C=(\frac{1}{2}(1-\tau),0),$ $D=(h_2^{-1}(1-h_2(\tau)),0)$.} 
    \label{fig:bound}
    \vspace{-20pt}
\end{figure}

Fig.~\ref{fig:bound} shows the lower bound on the rate from Theorem~\ref{th::random} compared to the upper bound (\ref{upper_lp}). We note that the gap between points A and B, C and D decreases with a decrease of $\tau$.

\bibliographystyle{IEEEtran}
\bibliography{main}

\end{document}